\documentclass[prb,onecolumn,showpacs,preprintnumbers,amsmath,amssymb]{revtex4}
\usepackage{graphicx}
\usepackage{dcolumn}
\usepackage{bm}

\begin{document}

\title{The interaction between a superconducting vortex and an out-of-plane magnetized ferromagnetic disk: influence of the magnet geometry}

\author{M. V. Milo\v{s}evi\'{c}}
\author{F. M. Peeters}
\email{peeters@uia.ua.ac.be}

\affiliation{Departement Natuurkunde, Universiteit Antwerpen (UIA), \\
Universiteitsplein 1, B-2610 Antwerpen, Belgium}

\date{\today}

\begin{abstract}
The interaction between a superconducting vortex in a type II
superconducting film (SC) and a ferromagnet (FM) with out-of-plane
magnetization is investigated theoretically within the London
approximation. The dependence of the interaction energy on the
FM-vortex distance, film thickness and different geometries of the
magnetic structures: disk, annulus(ring), square and triangle are
calculated. Analytic expressions and vectorplots of the current
induced in the SC due to the presence of the FM are presented. For
a FM disk with a cavity, we show that different local minima for
the vortex position are possible, enabling the system to be
suitable to act as a qubit. For FMs with sharp edges, like e.g.
for squares and triangles, the vortex prefers to enter its
equilibrium position along the corners of the magnet.
\end{abstract}
 \maketitle

\section{Introduction}

The interaction between superconductivity and magnetism has drawn
a lot of attention in the last decades. To study the effects due
to the interplay of the superconducting order parameter and the
non-homogeneous magnetic field resulting from a ferromagnet (FM),
several experimental groups fabricated periodic arrays of magnetic
dots and antidots positioned above or under the superconducting
film.~\cite{baert,schuller,leuven1,leuven2} Such ferromagnetic
dots act as very effective trapping centers for the vortices which
leads to an enhancement of the critical current. Recently, it was
predicted~\cite{bul} that an increase of the pinning effects by
two orders of magnitude can be realized in this way. After
substantial progress in the preparation of regular magnetic arrays
on superconductors and considering the importance of such
structures for magnetic device and storage technologies, these
hybrid systems became very interesting both from a theoretical and
an experimental point of view. Macroscopic phenomena have already
been explored experimentally, but a theoretical analysis of the
magnetic and superconducting response in such systems is still in
its infancy.

In previously proposed models for the superconducting film (SC)
interacting with a ferromagnet (FM) on top of
it~\cite{lip1,kay1,kay2,helseth} the magnetic texture interacts
with the SC current, which subsequently changes the magnetic
field. The authors used the London approximation to describe this
system since the sizes of all structures are much larger than the
coherence length $\xi$. The thickness of the SC film and of the FM
was assumed to be negligibly small (i.e. $d<<\xi,\lambda$).
Elementary solutions for the interaction between the circular
magnetic dot (bubble) or annulus (ring) with a vortex were found.
Further, the creation of additional vortices by the FM near the
surface was described~\cite{kay2}, by a simple comparison of the
free energies of the system with and without the vortex. However,
the spontaneous creation of a vortex-antivortex pair as a possible
lower energy state was never considered.

Other theoretical studies involving finite size ferromagnets were
mainly restricted to the problem of a magnetic dot with
out-of-plane magnetization embedded in a superconducting
film.~\cite{marm,fertig} Marmorkos \textit{et al.}~\cite{marm}
were the first to solve the non-linear Ginzburg-Landau (GL)
equation numerically, with appropriate boundary conditions for an
infinitely long ferromagnetic cylinder penetrating the
superconducting\ film, and found a correspondence between the
value of the magnetization and the vorticity of the most
energetically favorable giant-vortex state. The vortex structure
of a SC disk with a smaller magnetic disk on top of it was
numerically calculated in Ref.~\cite{misko1} using the first
non-linear GL equation, i.e. neglecting the effect of the
screening currents on the total magnetic field. Interesting
vortex-antivortex configurations and an interplay between the
giant-vortex and multi-vortices were found.

Most recently, the pinning of vortices by small magnetic particles
was studied experimentally~\cite{morgan,bael,mel} which was a
motivation for our recent theoretical study of this
system~\cite{misko2}. In the latter study we approximated the
magnetic field profile by a magnetic dipole. In the present paper,
we generalize these results in order to include the {\it realistic
magnetic field profile} of the FM which in the present approach
can be of arbitrary shape. The superconducting film lies in the
xy-plane while the FM is positioned a distance $l$ above the SC,
and is magnetized in the positive z-direction (out-of-plane). To
avoid the proximity effect and the exchange of electrons between
the FM and the SC we assume a thin layer of insulating oxide
between them as is usually the case in the experiment.

The paper is organized as follows. In the next section we present
the general formalism. In Sec.~III, we discuss the pinning
potential of the magnetic disk and magnetic annulus (ring) with
out-of-plane magnetization. Further, the vortex-magnet interaction
energy and supercurrent induced in the superconductor are
calculated analytically and the profiles are shown. We use these
results in Sec.~IV to investigate the manipulation of vortices in
the case of a more complicated geometry of the magnet, i.e.
magnetic disk with an off-center hole(s). In Sec.~V the pinning
properties of the magnet with square or triangular shape are
analyzed and the most favorable trajectory of the vortex with
respect to the magnet edge is determined. The influence of edges
(corners) of the FM on the pinning is then discussed and our
conclusions are given in Sec.~VI.

\section{Theoretical formalism}

We consider a ferromagnet of arbitrary shape with homogeneous
out-of-plane magnetization $\vec{M}$, placed outside a type II SC
film interacting with a single vortex in the SC. Within the London
approximation, the direct interaction energy between the vortex
and the FM in a stationary magnet-superconductor system is given
by~\cite{misko2}
\begin{equation}
U_{mv}=\frac{1}{2c}\int dV^{(i)}\left[ \vec{j}_{m}\cdot
\vec{\Phi}_{v}\right] -\frac{1}{2}\int dV^{(fm)}\vec{h}_{v}\cdot
\vec{M}\text{,} \label{freemv}
\end{equation}
where $\vec{\Phi}_{v}=\left( \Phi _{\rho },\Phi _{\varphi },\Phi
_{z}\right) =\left( 0,\Phi _{0}/(2\pi \rho ),0\right) $ denotes
the vortex magnetic flux vector ($\Phi _{0}$ is the flux quantum).
The first integration is performed over the volume inside
$V^{(i)}$ the superconductor, while $V^{(fm)}$ in the second
integral denotes the volume of the ferromagnet. Indices $v$ and
$m$ refer to the vortex and the magnet, respectively, $\vec{j}$
denotes the current and $\vec{h}$ the magnetic field.

The interaction energy in this system consists of two parts: (i)
the interaction between the Meissner currents generated in the SC
($\vec{j}_{m}$) by the FM and the vortex, and (ii) the interaction
between the vortex magnetic field and the FM. In
Ref.~\cite{misko2} we showed analytically that in the case of a
point magnetic dipole (MD) for both in- and out-of-plane
magnetization these two contributions are equal. Due to the
superposition principle, the finite FMs with homogeneous
magnetization can be represented as an infinite number of dipoles.
Consequently, in our case of out-of-plane magnetized FM, the
vortex-magnet interaction energy equals
\begin{equation}
U_{mv}= -\int dV^{(fm)}\vec{h}_{v}\cdot \vec{M}\text{.}
\label{freemv1}
\end{equation}

In order to obtain the current induced in the superconductor by
the ferromagnet, one should solve first the equation for the
vector potential~\cite{mel}
\begin{equation}
rot\left( rot\vec{A}_{m}\right) +\frac{1}{\lambda ^{2}}\Theta
\left( d/2-\left| z\right| \right) \vec{A}_{m}=4\pi rot\vec{M}.
\label{cur}
\end{equation}
This equation is rather complicated to be handled for a finite
size FM, but, the analytic expressions for the induced SC current
in an infinite superconducting film with thickness $d$
($-\frac{d}{2}<z<\frac{d}{2}$) in the MD case (with magnetic
moment $m$) are known~\cite{misko2}
\begin{subequations}
\label{curr}
\begin{eqnarray}
j_{x}^{md}(x,y,z)=\frac{cm\Phi _{0}}{2\pi \lambda ^{3}}
\frac{y_{m}-y}{R_{m}}\int\limits_{0}^{\infty }dq\exp \left\{
-q\left( \left| z_{m}\right| -\frac{d}{2} \right)
\right\}q^{2}J_{1}(qR_{m})C\left( q,z\right) \text{,} \label{c1}
\end{eqnarray}
\begin{eqnarray}
j_{y}^{md}(x,y,z)=\frac{cm\Phi _{0}}{2\pi \lambda ^{3}}
\frac{x-x_{m}}{R_{m}}\int\limits_{0}^{\infty }dq\exp \left\{
-q\left( \left| z_{m}\right| -\frac{d}{2} \right)
\right\}q^{2}J_{1}(qR_{m})C\left( q,z\right) \text{,} \label{c2}
\end{eqnarray}
\end{subequations}
with
\begin{equation}
C\left( q,z\right)
=\frac{k~\text{cosh}(k(\frac{d}{2}+z))+q~\text{sinh}(k(\frac{d}{2}+z))}{
(k^{2}+q^{2})~\text{sinh}(kd)+2kq~\text{cosh}(kd)}\text{,}
\label{c3}
\end{equation}
where $k=\sqrt{1+q^{2}}$, $R_{m}
=\sqrt{(x-x_{m})^{2}+(y-y_{m})^{2}}$ is the distance between the
dipole and the point of interest, and $J_{v}(\alpha)$ is the
Bessel function. The coordinates $(x_{m},y_{m},z_{m})$ denote the
position of the dipole. The magnetic moment of the magnet is
measured in units of $m_{0}=\Phi _{0}\lambda $, and all distances
are scaled in units of $\lambda$. These units will be used in the
rest of the paper.

To find the supercurrent induced by a finite size FM above the
superconductor, we make use of the superposition principle and
consequently the above expressions (\ref{curr},\ref{c3}) have to
be integrated over the volume of the ferromagnet. Thus, the value
of the current is given by ($\alpha=x,y$)
\begin{equation}
j_{\alpha}(x,y,z)=\int j_{\alpha}^{md}(x,y,z)~dV^{(fm)}\text{.}
\label{current}
\end{equation}

\section{Magnetic disk (ring) - vortex interaction}

In this section, we investigate the interaction between a vortex
in an infinite type II superconducting film with thickness $d$
($-\frac{d}{2}<z<\frac{d}{2}$) and the magnetic disk with radius
$R$ and thickness $D$ with \textit{out-of-plane magnetization},
i.e. $\overrightarrow{M}=\Theta (R-\rho)\Theta (z_{1}-z)\Theta
(z-z_{0})m\overrightarrow{e_{z}}/(R^2 \pi D)$ (in units of
$M_{0}=\Phi/\lambda^2$) located at distance $l$ above (under) the
SC ($z_{0}=l$, $z_{1}=l+D$).

Inserting the well known expression for the magnetic field of a
vortex outside the SC~\cite{misko2}
\begin{subequations}
\label{Hvr2}
\begin{equation}
h_{vz}\left( \rho ,z\right) =\frac{L\Phi _{0}}{2\pi \lambda
^{2}}\int\limits_{0}^{\infty }\frac{dq~q}{Q}J_{0}\left( qR
\right)\exp \left( -q \frac {2\left| z\right| -d}{2}\right),
\end{equation}
\begin{equation}
h_{v\rho }\left( \rho ,z\right) =\frac{L\Phi _{0}}{2\pi \lambda
^{2}} \mathop{\rm sgn} (z)\int\limits_{0}^{\infty
}\frac{dq~q}{Q}J_{1}\left( qR \right) \exp \left( -q \frac
{2\left| z\right| -d}{2} \right),
\end{equation}
\end{subequations}
into Eq.~(\ref{freemv1}) we find the expression for the magnetic
disk-vortex interaction
\begin{subequations}
\label{Udisk}
\begin{equation}
U_{mv}=\frac{MR\Phi _{0}^{2}}{\lambda}~U_{\perp }\left( \rho
_{v}\right) \text{,}
\end{equation}
where $\rho =\rho _{v}$ denotes the position of the vortex, and
\begin{equation}
U_{\perp }\left( \rho _{v}\right) =\int\limits_{0}^{\infty
}dq\frac{1}{qQ} J_{1}\left( qR\right) J_{0}\left( q\rho
_{v}\right) E(q,l,D) \text{,}
\end{equation}
\end{subequations}
where $Q=k(k+q\coth (kd/2))$, and $E\left( q,l,D \right) =
e^{-ql}\left(e^{-qD}-1\right)$.

For the case of a thin ferromagnetic disk above the thin
superconducting film ($d,D<1$), the following asymptotics can be
obtained:

\begin{subequations}
\label{asym}
(1) for $\rho _{v}<R,$ we found
\begin{equation}
U_{\perp }\left( \rho _{v}\right) \approx -\frac{dDR}{2}\left(
\frac{1}{l+ \sqrt{l^{2}+R^{2}}}-\frac{\rho
_{v}^{2}}{4}\frac{1}{\left( l^{2}+R^{2}\right) ^{3/2}}\right) ;
\end{equation}

(2) for $\rho _{v}>R$ and $\rho _{v} \sim l,$
\begin{equation}
U_{\perp }\left( \rho _{v}\right) \approx -\frac{dDR}{4}\left(
\frac{1}{\sqrt{l^{2}+\rho
_{v}^{2}}}-\frac{R^{2}}{8}\frac{2l^{2}-\rho _{v}^{2}}{\left(
l^{2}+\rho _{v}^{2}\right) ^{5/2}}\right) ;
\end{equation}

(3) for $\rho _{v}>R,l,$
\begin{eqnarray}
U_{\perp }\left( \rho _{v}\right) \approx -\frac{dDR}{4}\left(
\frac{1}{\sqrt{l^{2}+\rho _{v}^{2}}}-\frac{d\pi }{4}\left[
H_{0}\left( \frac{\rho _{v}d}{2}\right) -Y_{0}\left( \frac{\rho
_{v}d}{2}\right) \right] \right) .
\end{eqnarray}
\end{subequations}
Here, $H_{v}(x)$ and $Y_{v}(x)$ denote the Struve and Bessel
function, respectively.

In the case of a FM on top of the SC ($l=0$), the above
asymptotics (\ref{asym}(a-c)) can be expressed in a more precise
way:

\begin{subequations}
\label{asym1}
(1) for $\rho _{v}<R,$
\begin{equation}
U_{\perp }\left( \rho _{v}\right) \approx -\frac{dD}{\pi }E\left(
\frac{\rho _{v}^{2}}{R^{2}}\right) ;
\end{equation}

(2) for $\rho _{v}>R,$
\begin{eqnarray}
U_{\perp }\left( \rho _{v}\right) \approx -\frac{dDR}{\pi }\left\{
\frac{1}{\rho _{v}}\left[ \frac{\rho _{v}^{2}}{R^{2}}E\left(
\frac{R^{2}}{\rho _{v}^{2}}\right)+\left( 1-\frac{\rho
_{v}^{2}}{R^{2}}\right) K\left( \frac{R^{2}}{\rho _{v}^{2}}\right)
\right]+\frac{d\pi ^{2}}{16}\left[ H_{0}\left( \frac{\rho
_{v}d}{2}\right) -Y_{0}\left( \frac{\rho _{v}d}{2}\right) \right]
\right\} .
\end{eqnarray}
\end{subequations}
$K(x)$ and $E(x)$ are the complete elliptic integrals of the first
and second kind, respectively. Further expansion of the asymptotic
behavior of the energy at large distances (Eqs. (\ref{asym}c) and
(\ref{asym1}b)) gives $U_{\perp }\left( \rho _{v}\right) \approx
-DR\big/d\rho _{v}^{3}$.

When we take the derivative of the interaction energy,
Eq.~(\ref{Udisk}), over $\rho_{v}$ we obtain the force acting on a
vortex in the presence of a magnetic disk:
\begin{subequations}
\label{forcedisk}
\begin{equation}
F_{mv}=\frac{MR\Phi _{0}^{2}}{\lambda^2}~F_{\perp }\left( \rho
_{v}\right) \text{,}
\end{equation}
with
\begin{eqnarray}
F_{\perp }\left( \rho _{v}\right) =\int\limits_{0}^{\infty
}dq\frac{1}{Q} J_{1}\left( qR\right) J_{1}\left( q\rho _{v}\right)
E(q,l,D) \text{.}
\end{eqnarray}
\end{subequations}

For the case of a thin FM on top of a thin SC ($d,D<1,$ $l=0$), we
derived the following asymptotics

\begin{subequations}
(1) for $\rho _{v}<R,$
\begin{equation}
F_{\perp }\left( \rho _{v}\right) \approx \frac{dD}{\pi \rho
_{v}}\left[ E\left( \frac{\rho _{v}^{2}}{R^{2}}\right) -K\left(
\frac{\rho _{v}^{2}}{R^{2}}\right) \right] ;
\end{equation}

(2) for $\rho _{v}>R,$
\begin{eqnarray}
F_{\perp }\left( \rho _{v}\right) \approx \frac{dD}{\pi R}\left[
E\left( \frac{R^{2}}{\rho _{v}^{2}}\right) -K\left(
\frac{R^{2}}{\rho _{v}^{2}} \right)
\right]-\frac{d^{3}DR}{16}\frac{\pi }{2}\left[ Y_{1}\left(
\frac{\rho _{v}d}{2}\right) +H_{-1}\left( \frac{\rho
_{v}d}{2}\right) \right].
\end{eqnarray}
\end{subequations}
The latter expression reduces in the extreme $\rho _{v}>>R$ limit
to $F_{\perp }\left( \rho _{v}\right) \approx -3DR\big/d\rho
_{v}^{4}$, which is consistent with the asymptotic behavior of the
interaction energy.

The results for the full numerical calculation of Eq.
(\ref{Udisk}) are shown in Fig.~1(a) for a magnetic disk with
radius $R=3.0$, and three values of the thickness $D=0.1,0.5,1.0$,
fixed total magnetic moment $m=1.0$ ($M=m/V^{(fm)}$), and at
distance $l=0.1$ above the SC with thickness $d=0.1$. The energy
is expressed in units of $U_{0}=\Phi _{0}^{2}\big/\pi\lambda)$ and
the force in $F_{0}=\Phi _{0}^{2}\big/\pi\lambda^{2}$. The
magnet-vortex interaction energy increases if the magnet is made
thinner, since the magnetization in that case increases. Also the
magnetic field of the disk becomes more peaked near the magnet
edge. In Fig.~1(b), the dependence of the interaction energy on
the thickness of the SC is shown. The increased thickness of the
SC makes the interaction stronger, due to the stronger response of
the SC to the presence of the magnet. Notice that increasing the
thickness beyond $d>>\lambda$ does not influence the energy
(dashed curve in Fig.~1(b), $Q\approx k(k+q)$ in
Eq.~(\ref{Udisk})).

\begin{figure}[b]
\caption{\label{fig:fig1}The magnetic disk-vortex interaction
energy as function of the distance between the vortex and the
center of the FM disk: for several values of (a) the thickness of
the FM disk, and (b) the thickness of the SC. (c) Plot of the
FM-vortex force, and (d) a vectorplot of the current induced in
the SC due to the presence of the FM (same parameters as in (c)).
The grey semicircle in (d) indicates the position of the edge of
the FM.}
\end{figure}

The vortex is attracted by the magnetic disk when the
magnetization and the vortex are oriented parallel, independently
of the value of the parameters. The interaction energy has its
minimum just under the center of the disk, which is the
energetically most favorable position of the vortex. The force
acting on the vortex is purely attractive and it has its maximum
at the edge of the magnetic disk (see Fig.~1(c)). For large
distances between the magnetic disk and the vortex the interaction
approaches the value obtained earlier for the case of a magnetic
dipole.\cite{misko2} Note that in the limit $R\rightarrow0$ and
$D\rightarrow0$, Eq.~(\ref{Udisk}) corresponds to the out-of-plane
dipole case of Ref.~\cite{misko2}. In Refs.~\cite{kay2,helseth}
the interaction energy between a magnetic nanostructure and a
vortex in a thin superconductor ($d<<1$) was calculated. In their
case, the thickness of the magnet was not taken into account
(assumed to be infinitely thin), the superconducting film was
taken very thin ($d<<\lambda$) and the FM was placed on top of the
SC (in the same $z=0$ plane) which corresponds to $l=0$ in our
case. In these limits, our equations reduce to the same
expressions for the interaction energy like those given in
Refs.~\cite{kay2,helseth}. But note that the analytical
expressions are still not completely reached in the distance range
shown in Fig.~1 and therefore a numerical calculation of the full
integral is necessary in order to obtain the magnet-vortex
interaction energy. Therefore, from this point of view, our
expression offers much more information (non-zero thickness of
both FM and SC, and arbitrary position of the FM above the SC)
without a real increase of the complexity of the numerical
calculation.

To better understand the attractive magnet-vortex behavior in this
system, we calculated the supercurrent induced in the SC due to
the presence of the magnet. As explained in Sec. II, this current
can be obtained after integration of Eq.~(\ref{current}). In the
case of a flat magnetic disk, it has only an azimuthal component
and reads
\begin{equation}
j_{\varphi }(\rho ,z)=\frac{cMR\Phi _{0}}{\lambda
^{3}}\int\limits_{0}^{\infty }dqJ_{1}(qR)J_{1}(q\rho)
E\left(q,l,D\right) C\left( q,z\right) \text{,} \label{curdisk}
\end{equation}
where $\rho =\sqrt{x^{2}+y^{2}}$, and $C(q,z)$ is given by
Eq.~(\ref{c3}). For a FM placed under the SC, one should replace
$z$ by $-z$. The vectorplot of the current is shown in Fig.~1(d).
One should notice that the direction of the current is the one
normally associated with an antivortex (the clockwise direction)
and that the current is maximal at the magnetic disk edge. This
agrees with our previous results: the direction of the current
explains the attraction between the FM disk and the vortex, and
the position of the maximum of the current corresponds to the
maximal attractive force. The problem is cylindrically symmetric,
and a vortex approaching the magnet from any direction will be
attracted for parallel alignment and repelled in the anti-parallel
case. This important point was not fully explained in
Ref.~\cite{helseth}.

Using the same procedure, for a magnetic annulus $(R_{i}<\rho
<R_{o})$ with thickness $D$ and out-of-plane magnetization (inset
in Fig.~2(b)) we have
\begin{equation}
\overrightarrow{M}=\frac{1}{V_{ann}}\Theta (\rho-R_{i})\Theta
(R_{o}-\rho)\Theta (z_{1}-z)\Theta
(z-z_{0})m\overrightarrow{e_{z}},
\end{equation}
resulting in the vortex-magnet interaction energy
\begin{eqnarray}
U_{mv} =\frac{M\Phi _{0}^{2}}{\lambda} \int\limits_{0}^{\infty
}dq\frac{1}{qQ}J_{0}\left( q\rho_{v}\right) \left(
R_{o}J_{1}(qR_{o})-R_{i}J_{1}(qR_{i})\right)E(q,l,D)\text{.}\label{Uann}
\end{eqnarray}

The interaction energy and force curves for the magnetic
annulus-vortex interaction are given in Fig.~2(a,b) and are in
qualitative (but not quantitative) agreement with earlier results
of Ref. \cite{kay1}, which were obtained in the limit of an
extremely thin SC and FM, namely $d,D<<\lambda $. Please notice
again that in our calculation finite thicknesses of both SC and FM
are fully taken into account. The most important result is that,
in this case, the annulus-vortex interaction energy has a
ring-like minimum, under the magnet. The exact radial position of
this minimum depends on the SC parameters, the thickness of the
magnet and its distance from the SC. The force acting on the
vortex shows dual behavior - attractive outside the equilibrium
ring and repulsive inside.

\begin{figure}[h]
\caption{\label{fig:fig2}The out-of-plane
magnetized annulus-vortex interaction: (a) the interaction energy,
and (b) plot of the FM-vortex force. The contourplot of the
interaction energy is shown as inset in (a) (dark color
illustrates low energy, as will be the case throughout the paper,
the dashed white semicircles illustrate the edges of the magnet).
A schematic outline of the magnetic annulus is shown in the inset
of (b).}
\end{figure}

Due to the dual behavior of the FM-vortex force one naively
expects different current flow in the superconductor inside and
outside the annulus. We use again Eq.~(\ref{current}) and for the
current induced in the superconductor we obtain
\begin{eqnarray}
j_{\varphi }(\rho ,z) =\frac{cM\Phi _{0}}{\lambda
^{3}}\int\limits_{0}^{\infty }dq\left(
R_{o}J_{1}(qR_{o})-R_{i}J_{1}(qR_{i})\right)J_{1}(q\rho ) E(q,l,D)
C\left( q,z\right).\label{jann}
\end{eqnarray}
In Fig.~3 we show a comparison between the currents induced in the
SC in the case of a magnetic disk (dashed curves) and a magnetic
annulus (solid curves), for different SC-FM vertical distances.
When the magnet is positioned far above the superconductor, the
vortex qualitatively does not feel the difference between the disk
and the ring case, and the current induced in the SC shows a
similar behavior (Fig.~3(a)). When approaching the superconductor,
the influence of the central hole in the ring becomes more
pronounced (Fig.~3(b,c)), and eventually the current changes sign
(Fig.~3(d)).

\begin{figure}[b]
\caption{\label{fig:fig3}Comparison between the current induced in
a SC (in units of $j_{0}=c\Phi _{0}\big/\pi\lambda ^{3}$) by a
magnetic disk (dashed curves) and a magnetic annulus (solid
curves) with the same outer radius, as the FM-SC vertical distance
$l$ decreases (a)-(d).}
\end{figure}

Obviously, the qualitative behaviors of all quantities outside the
annulus approach those for the case of a magnetic disk. However,
inside the ring, the situation is different. The nature of the
magnet-vortex force changes and while the current flows in the
clockwise direction outside the ring, inside the superconductor
the direction of the current is anti-clockwise in the case of a
small FM-SC distance (i.e. $l/\lambda=0.1$, Fig.~3(d)). Please
notice that due to the fact that the finite thickness of the SC is
included in our calculations, the SC current contains also a
z-dependence (i.e. through the $C(q,z)$ function).

From a look at Eqs. (\ref{Uann}) and (\ref{jann}), one can see
that the problem of a magnetic annulus actually can be modelled by
two concentric magnetic disks with different radius and opposite
magnetization. The problem is linear, and this will facilitate the
calculation in the cases of non-cylindrically symmetric FMs.

\section{Manipulation of the equilibrium vortex position with a magnetic disk containing a cavity}

In previous section, we discussed the pinning of vortices by a
magnetic disk or annulus (ring). We showed that the most
energetically favorable position for the vortex is under the
center of the magnetic disk (for parallel alignment) or under the
annulus (equilibrium ring). Here we generalize the latter system
and displace the hole in the disk from its central location.

Referring to the previous section, we may consider this problem as
a superposition of effects of two magnetic disks with opposite
magnetization. The smaller radius magnetic disk with the opposite
magnetization models the hole in the larger disk. The parameters
of the magnet are: the outer radius $R_{o}$, the radius of the
hole $R_{i}$, the center of the hole is at $(x_{h},y_{h})$ and the
thickness of the FM is denoted by D. Therefore, using
Eqs.~(\ref{Udisk}-\ref{curdisk}) for two magnetic disks, one with
radius $R_{o}$, and the other with radius $R_{i}$, with opposite
magnetization and centered at $(x,y)=(0,0)$ and $(x_{h},y_{h})$,
respectively, we investigate the pinning properties of such FM.

The results of this calculation are shown in Fig.~4 for a magnetic
disk with $R_{o}=3.0$, $R_{i}=1.0$, $(x_{h},y_{h})=(0.5,0)$,
$l=0.1$ and $D=d=0.5$. In Fig.~4(a) we show the contourplot of the
FM-vortex attractive force. It is clear that there are two local
energy minima along the $y=0$ direction, where the force equals
zero: in front of and behind the hole. In Fig.~4(b) the plot of
the interaction energy is given along this direction for three
positions of the hole $x_{h}=0.5,0.8$, and $1.0$. The important
result is that the two minima are not equal: the one closer to the
outer edge of the magnet has higher energy (metastable state) and
the one near the magnet center is the actual ground
state.
\begin{figure}[h]
\caption{\label{fig:fig4}The interaction
of the vortex with a magnetic disk with an off-center hole (see
left figure for schematics of the magnet configuration): (a)
contourplot of the interaction energy for $R_{o}=3.0$,
$R_{i}=1.0$, $(x_{h},y_{h})=(0.5,0)$, $l=0.1$ and $D=d=0.5$, with
$M=M_{0}$ (dashed circles indicate the edges of the magnet), and
(b) the energy along the $x$-axis, for three different locations
of the off-center hole in the magnetic disk.}
\end{figure}
However, due to the presence of the hole, the equilibrium position
of a vortex is not exactly in the center, and depends on the
position of the hole. The magnet is not cylindrically symmetric
and we have two separate energy minima instead of a ring of minima
as in the case of the magnetic annulus. Also, one could argue that
a slowly moving vortex in a system with no temperature
fluctuations could be trapped at the metastable position. Anyhow,
the hole in a magnetic disk appears to be a powerful tool for a
possible manipulation of the vortex position. However, one
question arises: since there are two minima present in the
interaction energy, is it possible to have two equilibrium states
with the same energy?

In order to construct such a situation, we introduced a second
hole in the magnet, at a symmetrical position to the first one
with respect to the center of the magnet. As an example, we took
the parameters of the magnet as $R_{o}=3.0$,
$R_{i}(for~both~holes)=0.5$, $(x_{h},y_{h})=(\pm1.0,0)$. The
interaction energy along the $x$-axis is given in Fig.~5. Two
equal minima near the outer edge of the disk are found next to the
magnet holes. However, the global minimum is still under the
center of the magnet (see inset of Fig.~5).

\begin{figure}[b]
\caption{\label{fig:fig5}The magnetic disk with two symmetrical
holes: plot of the FM-vortex interaction energy illustrating the
position of the metastable and ground vortex states with respect
to the position of the holes (dashed vertical lines). The
parameters of the system are $R_{o}=3.0$,
$R_{i}(for~both~holes)=0.5$, $(x_{h},y_{h})=(\pm1.0,0)$, $l=0.1$,
$D=d=0.5$ and $M=M_{0}$.}
\end{figure}

\begin{figure}[h]
\caption{\label{fig:fig6}The interaction energy of a vortex with a
magnetic disk containing an ``eight-hole'': (a) interaction energy
for the vortex positioned along the $x$-axis at the $y=0$ line,
(b) along the $y$-axis for $x=0$, and (c) contourplot of the
FM-vortex interaction energy (dark color-low energy). In (c) the
vectorplot of the current induced in the SC in the presence of the
FM is superimposed. The solid circles denote the position of the
holes in the FM above the SC.}
\end{figure}

To eliminate this minimum we allow the holes to touch each other
and to form an ``eight-hole'' in the center of the magnet. The
resulting interaction energy (Fig.~6(a)) has now only two equal
minima along the $y=0$ direction, outside the hole, near the
magnet edge, and a plateau-like behavior in the central region.
Still, these minima are not the lowest energy states. The central
global minimum of the interaction energy from the previous case is
now split by the joined holes into two minima along the $x=0$
direction (see Fig.~6(b)). The latter minima represent the ground
state for a vortex in the presence of a magnetic disk with an
``eight-hole''. The vortex has two absolutely equal ground states
and the same probability of eventually sitting in one of those.
Thanks to this feature, a possible use of this system for quantum
computing can be analyzed, similarly to the quantum systems
proposed before (see, for example, Ref.~\cite{denis}).

In Fig.~6(c) the contourplot of the magnet-vortex interaction
energy is given, together with a vectorplot of the current induced
in the SC. The position of the eight-hole is denoted by the thick
solid circles. Around the magnetic disk, the SC current flows in a
clockwise direction, illustrating the general attraction between
the FM and the vortex. However, under the magnet, current shows a
dual behavior, and a vortex-antivortex-like current flow can be
seen. Namely, at the equilibrium vortex-states we find
``antivortex'' current profiles while under the holes of the
magnet a vortex-like current motion is present. This suggests the
possibility that such a magnetic field configuration, for
sufficiently strong magnetization, could induce interesting
vortex-antivortex configurations if placed near a superconducting
film.

Using this approach, interaction of a vortex with magnets of more
complicated shapes can be investigated. We have shown that the
magnetic disk with a cavity is a nice example of how to control
the vortex by the magnet-geometry. From the point of view of
practical vortex manipulation, it would be interesting to move the
vortex by changing the parameters of the system.
Helseth~\cite{helseth} proposed a system in which a magnetic disk
(with magnetization $M_{1}$) is placed in the center of a magnetic
ring (with magnetization $M_{2}$), where the disk and the ring can
have opposite magnetization. In Fig.~7 we show the outline of the
system (upper inset), and the calculated interaction energy and
force acting on the vortex, for the parameters
$W_{1}=0.5\lambda_{e}$, $W_{2}=0.75\lambda_{e}$ and
$W_{3}=1.0\lambda_{e}$ and $D=0.1\lambda$ on top of a SC with
thickness $d=0.1\lambda$. Here $\lambda_{e}=\lambda^2/d$ denotes
the effective penetration depth. We suppose a thin oxide layer
between the SC and the FM with thickness $0.01\lambda$. First we
consider the $M_{1}=M_{2}$ case. Helseth claimed that in this
case, a slowly moving vortex will be attracted and sit under the
annulus. The plot of the interaction energy in Fig.~7(a) shows two
energy minima, namely, under the disk, and under the annulus
(ring-energy minimum, see inset of Fig.~7(a)).
\begin{figure*}[h]
\caption{\label{fig:fig7} Disk-ring magnetic structure (see inset
of (a) for a schematic view of the configuration) above the SC:
(a) FM-vortex interaction energy for $M_{1}=M_{2}$, (b) for
$M_{2}=-M_{1}$. The contourplots of the energy are given as insets
(dashed lines indicate the edges of the magnet). The force acting
on the vortex is shown in (c), as the solid curve for case (a) and
dashed one for case (b). The parameters are $W_{1}=0.5$,
$W_{2}=0.75$, $W_{3}=1.0$, $D=1.0$, and $l=0.1$, all in units of
the effective penetration depth $\lambda_{e}=\lambda^2/d$.}
\end{figure*}
From our
calculations it is clear that the ground state for the vortex
would be in the center of the magnetic structure and not under the
annulus as claimed in Ref.~\cite{helseth}. However, one could
argue that a vortex, slowly moving towards the magnet, could rest
in a metastable state under the ring, if there are no fluctuations
in the system. In Fig.~7(c), the solid line shows the force acting
on the vortex. Please note that in our calculation, the positions
of the extremes correspond to the magnet edges, while in
Ref.~\cite{helseth} this was not the case. Also, the peaks in the
forces in our calculation are finite, due to the finite thickness
of the magnet. It should be noted that the relation between the
two energy minima and the acting forces strongly depends on the
parameters and it can change in favor of the annulus if it is made
wider.

In Ref.~\cite{helseth} it is also stated that in the opposite
case, when the annulus magnetization changes sign, i.e.
$M_{2}=-M_{1}$, the vortex is attracted to the center of the disk.
The interaction energy we calculated in this case is plotted in
Fig.~7(b) (the force is given in Fig.~7(c) by the dashed line). It
is clear that the annulus forms an energetic barrier which
prevents a ``slowly moving'' vortex from reaching the central
position. However, the global minimum of the energy is under the
center of this magnetic structure (the energy shown in Fig.~7(b)
decreases monotonously and reaches zero at infinity), implying
that the vortex would definitely sit under the magnet for the
considered parameters. Different values for $R_{1},~R_{2},~R_{3}$
could make the central minimum higher, and the energy barrier
would then be able to repel the vortex. To conclude, in order to
use this magnetic structure for vortex manipulation, one should
not only overcome the experimental difficulties to realize such a
structure, but also be careful about the influence of the
parameters on the behavior of this system.

\section{Square and triangular magnetic disks: influence of the corners on the vortex pinning}

Up to this point, we have only considered the interaction of a
vortex with magnets having circular symmetry, namely magnetic
disks, rings, and combinations of those. Here we consider FMs with
broken circular symmetry, namely square or triangular magnetic
disks. We put the center of our Cartesian coordinates in the
center of the magnet and the $x$ axis along one of its sides. In
this case, Eq.~(\ref{freemv1}) can not be solved analytically, and
triple numerical integrations must be performed. In the case of a
rectangular FM we have
\begin{eqnarray}
U_{mv}(x_{v},y_{v}) =\frac{M\Phi
_{0}^2}{2\pi\lambda}\int\limits_{0}^{\infty
}\frac{dq}{Q}~E(q,l,D)\int\limits_{-A/2}^{A/2}dx\int\limits_{-B/2}^{B/2}dy~J_{0}\left(qR_{v}\right),
\end{eqnarray}
where $R_{v}=\sqrt{(x-x_{v})^2+(y-y_{v})^2}$. $A$ and $B$ are the
dimensions of the magnet (with thickness $D$) in the $x$ and $y$
direction, respectively. As before, the distance between the FM
and the SC surface is $l$.

The components of the current induced in the SC are obtained in an
analogous way, by numerical integration of Eq.~(\ref{current}),
using the expressions (\ref{c1}-\ref{c3}).

In Fig.~8(a) the interaction energy with a vortex is shown, along
two directions (lines of symmetry): (i) diagonal (dotted line);
(ii) horizontal (solid line). In the inset, the contourplot of the
energy is given. For comparison, we also give the results for a
magnetic disk (thick dashed curve) with $R^2\pi=AB$ and the same
thickness $D$. As far as the pinning potential of a square
magnetic disk is concerned, the asymmetry is rather small and the
result is not much different from the one of an equivalent
circular disk. Only in the region near the edge of the magnet,
some discrepancy between the pinning potential of the square
magnet in the diagonal direction and the disk approximation is
observed. Moving further from the magnet, this discrepancy
disappears. The energetically favorable position of the vortex is
under the center of the magnet. The situation is somewhat
different for rectangular-shaped FM disks. In Fig.~8(b) we show
the results for $A=4B=2.0$ and, in this case, the corresponding
circular disk becomes a very poor approximation. Far away from the
magnet, this approximation becomes better, as expected.

\begin{figure}[h]
\caption{\label{fig:fig8} Rectangular magnetic disk above a SC
film: (a) FM-vortex interaction energy for a square magnet case,
namely with sides $A=B=1.0\lambda$, thickness $D=0.1\lambda$ and
magnetization $M=M_{0}$, at $l=0.1$ above the SC with thickness
$d=0.1\lambda$ (the inset shows the contourplot of the energy,
white line indicates the edge of the magnet), and (b) for
$A=4B=2.0$. For comparison, the FM-vortex interaction energy in
the case of the magnetic disk with the same volume is given by the
thick dashed curves.}
\end{figure}

\begin{figure}[b]
\caption{\label{fig:fig9} The contourplots of the FM-vortex
attractive force for parallel orientation of the magnetization and
the vortex magnetic field, for: (a) an equilateral triangular
magnetic disk, with sides $a=1.0\lambda$, (b) square magnet, with
sides $A=B=1.0\lambda$. The other parameters correspond to the
ones in Fig.~8. The vectorplots of the current are superimposed.
In (c), the contourplot of the force for the rectangular magnet
case is given, $A=1.6\lambda$, $B=0.6\lambda$. Dark color
represents high force intensity. Positions of the edges of the
magnets are given by white solid lines.}
\end{figure}

From the behavior of the interaction energy we have seen that the
vortex is attracted to the center of the square or triangular
magnet for parallel orientation of the magnetization and the
vortex magnetic field. This corresponds qualitatively to the case
of a magnetic disk. However, the broken circular symmetry of the
magnet introduces some changes in the magnet-vortex interaction.
In Fig.~9 we show the contourplot of the force acting on the
vortex and the vectorplot of the current induced in the SC, both
for the case of a square and triangular magnetic disks. It is
obvious that the attractive force is stronger at the sides than at
the corners of the magnet. Therefore, the vortex approaching the
magnet at the side of the magnet will be attracted stronger than
on the diagonal direction. As far as the current is concerned, it
has the direction associated with an antivortex. Near the magnet,
the current follows the shape of the magnet and is maximal along
the sides of the magnet. Further from the square or triangular
magnet, the behavior of the current is more similar to the case of
the circular magnetic disk. In Fig.~9(c) the rectangular magnet
case is shown. One important feature should be noticed: the
FM-vortex attraction force is stronger at the longer side of the
rectangle.

As one can see in Fig.~9, the maxima of the FM-vortex interaction
force are located on the sides of the magnet. Therefore, one may
expect that the energetically preferable direction of vortex
motion when attracted by the FM is perpendicular to its sides. To
investigate this, we put the vortex in different positions (open
dots in Fig.~10(a)) and follow its trajectory using molecular
dynamics simulations. The results are shown in Fig.~10(a). If the
initial position of the vortex is along the lines of symmetry of
the magnet, the vortex moves straight towards the center of the
magnet. Otherwise, the trajectory of the vortex is distorted
towards the corner of the magnet. It actually appears that the
vortex {\bf avoids} the maxima of the attractive force. This is
counterintuitive but can be explained by the profile of the
FM-vortex interaction energy, given in Fig.~10(b,c). One should
notice the ``wave'' shape of the energy (if going along the ring
around the magnet, see fig.~10(b)). Following a circle around the
magnet, the energy has its minima at the corners of the magnet
(denoted by black triangles in Fig.~10(b,c)) and the saddle points
are on the sides (white triangles). The periodicity in Fig.~10(c)
corresponds to the number of corners of the ferromagnetic disk.

From Fig.~10 it is obvious that the interaction energy for any
position of the vortex lowers steep towards the center of the
magnet, but also towards the corners. This induces the distortion
of the vortex trajectory and gives the impression that the vortex
approaching the magnet from the corners is more energetically
favorable.

\begin{figure}[h]
\caption{\label{fig:fig10} The trajectory of the vortex when
attracted by the triangular magnetic disk (same parameters as in
Fig.~9(a)): (a) vortex paths with respect to the attractive force
landscape (the edge of the magnet is illustrated by white line),
(b) the contourplot of the triangular magnet-vortex interaction
energy (dark color - low energy), and (c) the interaction energy
along the ring indicated by dashed line in (b).}
\end{figure}

\section{Conclusion}

We applied the London theory to investigate flux pinning in SC
films due to the presence of a ferromagnet situated above (or
under) the SC, where the finite thickness of both FM and the SC
were fully taken into account. In the case of a magnetic disk or
annulus (ring), we obtained analytic expressions for the FM-vortex
interaction energy, force and the screening currents. We also
derived the asymptotic behavior of the interaction potential and
the force for specific values of the involved parameters. In the
case of a magnetic disk with an off-center hole we showed the
existence of two local minima in the FM-vortex interaction energy
- the ground state and the metastable one. By changing the
position of the hole, the position of the equilibrium moves with
respect to the magnetic disk center. We also showed that in the
case of a FM disk with two touching holes (``eight-hole'') two
minima with equal energy but different vortex position appeared.
The probability of a vortex sitting in one of these two states is
the same, which makes this system interesting as a possible qubit.
To further investigate the influence of the magnet geometry on its
pinning properties, we calculated the pinning potential for square
and triangular shaped ferromagnets. A substantial breaking of the
circular symmetry occurs and the attractive force acting on the
vortex is stronger at the sides of the magnet than at the corners.
Also, making one side of the rectangular magnet longer enlarges
the attractive force along it with respect to the other side.
Although counterintuitive, we showed that the vortex approaches
the non-circular magnet rather along the corners than
perpendicular to the sides, following the gradient of the
potential.

\section*{ACKNOWLEDGMENTS}

This work was supported by the Flemish Science Foundation (FWO-Vl), the
Belgian Inter-University Attraction Poles (IUAP), the ``Onderzoeksraad van
de Universiteit Antwerpen'' (GOA), and the ESF programme on ``Vortex
matter''. Stimulating discussions with D.~Vodolazov, V.~V.~Moshchalkov, and
M.~Van~Bael are gratefully acknowledged.

\end{document}